\documentclass[aps,prl,floats,twocolumn,superscriptaddress,showpacs]{revtex4}

\usepackage{graphicx,epsfig}% Include figure files
\usepackage{times}
\usepackage{graphics,dcolumn,bm,fleqn,epic,eepic,float}
\usepackage{amssymb,amsmath,multirow,rotate,color}
\begin{document}

\title{Enhancement of cooperation in highly clustered scale-free networks.}

\author{Salvatore Assenza}

\affiliation{Scuola Superiore di Catania, Via S. Paolo 73,
95123 Catania, Italy}

\author{Jes\'us G\'omez-Garde\~nes}

\affiliation{Scuola Superiore di Catania, Via S. Paolo 73,
95123 Catania, Italy}

\affiliation{Institute for Biocomputation and Physics of Complex
Systems (BIFI), University of Zaragoza, Zaragoza 50009, Spain}

\author{ Vito Latora}

\affiliation{Dipartimento di Fisica e Astronomia, Universit\`a di Catania and
INFN, Via S. Sofia 64, 95123 Catania, Italy}

\date{\today}

\begin{abstract}
We study the effect of clustering on the organization of cooperation, 
by analyzing the evolutionary dynamics of the Prisoner's Dilemma  
on scale-free networks with a tunable value of clustering. 
We find that a high value of the clustering coefficient produces an overall 
enhancement of cooperation in the network, even for a very high temptation to 
defect. On the other hand, high clustering homogeneizes the process of invasion  
of degree classes by defectors, decreasing the chances of survival of low densities of 
cooperator strategists in the network. 
\end{abstract}

 \maketitle

%%%%%%%%%%%%%%%%%%%%%%%%%%%%
%% INTRODUCTION
%%%%%%%%%%%%%%%%%%%%%%%%%%%%

%General
Cooperative phenomena are essential in natural and human systems and have been
the subject of intense research during decades
\cite{h64,ah81,wm00,ff03,ns05,n06}. Evolutionary game theory is concerned with
systems of replicating agents programmed to use some strategy in their
interactions with other agents, which ultimately yields a feedback loop that
drives the evolution of the strategies composition of the population
\cite{n06,hs98,hs03}. To understand the observed survival of cooperation among
unrelated individuals in populations when selfish actions provide a short-term
higher benefit, a lot of attention has been paid to the analysis of
evolutionary dynamics of the {\em Prisoner's Dilemma} (PD) game. In this simple
two-players game, individuals adopt one of the two available strategies,
cooperation (C) or defection (D); both receive $R$ under mutual cooperation
and $P$ under mutual defection, while a cooperator receives $S$ when
confronted to a defector, which in turn receives $T$, where $T>R>P>S$. Under
these conditions in a one-shot game it is better to defect, regardless of the
opponent strategy, and the proportion of cooperators asymptotically vanishes
in a well-mixed population.  On the other hand, the structure of interactions
among individuals in real societies are seen to be described by complex
networks of contacts rather than by a set of agents connected all-to-all
\cite{n03,blmch06}. Therefore, it is necessary to abandon the panmixia
hypothesis to study how cooperative behavior appear in the social context.

%% Previous studies
Several studies \cite{szabo,nm92,sp05,ak01,gcfm07,julia,ohln06,ezcs05,spl06}
have reported the asymptotic survival of cooperation on different kinds of
networks. Notably, cooperation even dominates over defection in  
non-homogeneous, {\em scale-free} (SF) networks, i.e. in graphs where   
the number $k$ of neighbors of an individual (the node degree) is 
distributed as a power law \cite{sp05,gcfm07}, $P(k)\sim k^{-\gamma}$, 
with $ 2<\gamma \le 3$.  
Networks with such a distribution are ubiquitous: scale-free topologies 
appear as the backbone of many social, biological, technological complex 
systems. However, in the context of social systems, other topological features,  
such as the presence of degree-degree correlations and of high clustering 
coefficients, are relevant ingredients to take into account in a complete 
description of the networks. The studies of the PD game on SF networks 
have considered so far networks with no degree correlations and nearly zero 
clustering coefficient, with the remarkable exception of Ref. 
\cite{PRSB} where high clustering SF networks are studied. Therefore, it is necessary 
to explore the effects that structural properties such as clustering and degre-degree correlations  
have on the survival of cooperation in complex netwoks.

%% Our goal
In this paper, we focus on the effects that the presence of non vanishing
clustering coefficient have on the survival of cooperation. The clustering 
coefficient of a network is related to the number of triangles present in 
the network, and is defined as the probability that two neighbors of a
given node share also a connection between them \cite{n03,blmch06}. 
A high clustering coefficient points out the presence of local neighborhoods, {\em i.e.} small
clusters of densely interconnected nodes, in the network. This property is
present in most of social networks where two friends of an individual are also
friends with high probability. Therefore a full description of cooperative
phenomena in social networks should be tackled by considering highly clustered
scale-free networks.

%% Networks
{\em Network model.-} We study a class of SF networks with
a tunable clustering coefficient introduced by Holme and Kim (HK) 
in Ref.~\cite{Holme}. The networks are constructed via a growing process that starts 
from an initial core of $m_{0}$ unconnected nodes. At each time step, a new node $i$ 
($i=m_0+1, ..., N$) is added to the network and links to $m$ (with $m\leq m_0$) of the 
previously existent nodes. The first link follows a preferential attachment 
rule (PA), {\em i.e.} the probability that node $i$ 
attaches to a node $j$  of the network (with $j<i$) is 
proportional to the degree $k_{j}$ of the node $j$. The remaining $m-1$ links
are attached in two different ways: {\em (i)} with probability $p$ 
the new node $i$ is connected to a randomly chosen neighbor of node $j$; 
{\em (ii)} with probability $(1-p)$ the PA rule is used again, and node $i$ is 
connected to another one of the previously existent nodes. 
With such a procedure one obtains SF networks with degree distribution $P(k)\sim k^{-3}$, 
and a tunable clustering coefficient depending on the value of $p$. 
In particular, for $p=0$ we recover the 
Barab\'asi-Albert model \cite{BA} where the clustering coeffienct tends to zero as the
network size $N$ goes to infinity. For values of $p>0$ the clustering coefficient
monotonously grows with $p$ \cite{Holme}. 

\begin{figure} %%%%%%%%%%%%%%  FIG 1
\epsfig{file=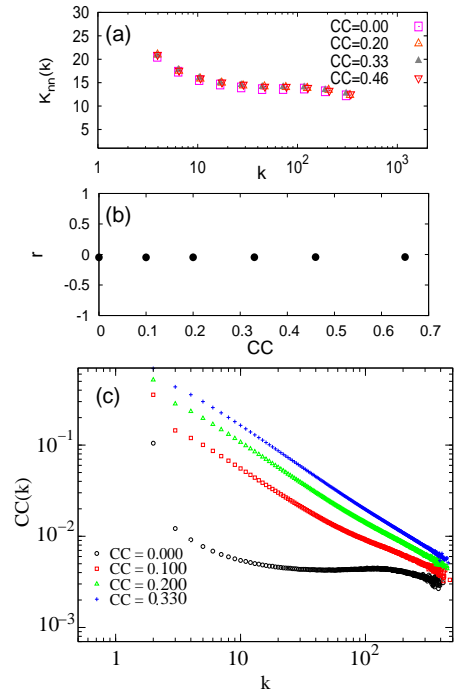,width=2.3in,angle=-0,clip=1}
\caption{(Color online). {\bf (a)} Average degree of the neighbors of nodes with degree 
$k$, $ K_{nn} (k)$, for four SF networks with different values of the $CC$.  {\bf (b)} Assortative 
index , $r$, as a function of the $CC$ of the networks. Both measures, $ K_{nn} (k)$ and 
$r$, reveal that SF networks generated from the HK model show no degree-degree 
correlations. {\bf (c)} Mean clustering coefficient of nodes with degree $k$, $CC (k)$, for 
four SF networks with different $CC$. From this figure it is clear a power 
law decay, $CC(k)\sim k^{-\alpha}$, for highly clustered networks.}
\label{fig:1}
\end{figure}

We have first checked that the networks produced by the HK model have
no degree-degree correlations, and we have analyzed the dependence of
the node clustering coefficient on the node degree. 
The clustering coefficient of a node $i$, $CC_i$,  
expresses how likely $a_{jm}=1$ for two neighbors $j$ and $m$ of
node $i$, where $A=\{a_{ij}\}$ is the adjacency matrix of the graph. 
The value of $CC_i$ is obtained by counting the actual number of
edges, denoted by $e_{i}$, in $G_i$, the subgraph of neighbors of
$i$. The clustering coefficient of $i$  is
defined as the ratio between $e_i$ and ${k_i(k_i-1)}/2$, the
maximum possible number of edges in $G_i$ \cite{n03,blmch06}:  
\begin{equation} \label{ci}
  CC_i=\frac{2e_i}{k_i(k_i-1)}=
   \frac{\sum_{j,m}a_{ij}a_{jm}a_{mi}}{k_i(k_i-1)} \ .
\end{equation} 
The mean clustering coefficient of the graph, $CC$, is then
given by the average of $CC_i$ over all the nodes in the network.
By definition, $0 \leq CC_i \leq 1$ and $0 \leq CC \leq 1$.  In
Fig.~\ref{fig:1} we report the results obtained for networks with
$m=m_{0}=3$ and $N=5\cdot 10^3$.  We have considered different values
of $p$ corresponding to networks with mean clustering coefficient 
$CC= 0, 0.1, 0.2, 0.33, 0.46$ and $0.65$. 
Ensembles of $2\cdot 10^4$ networks have been generated for
each value of $p$.  In Fig.~\ref{fig:1}.a we plot, as a function of
$k$, the average degree $K_{nn} (k) $ of the neighbors
of nodes with degree $k$. The figure shows a nearly constant function
$K_{nn} (k)$, pointing out that the HK model produces SF networks with
no degree-degree correlations. This result is further confirmed  by computing 
the assortative index $r$, introduced in Ref. \cite{Newman-assort}, as a punction of the networks' $CC$. As observed from Fig. \ref{fig:1}.b the values of $r$ are close to $0$ for all values of the $CC$, thus confirming the absence of degree-degree correlations in all the studied networks. 
On the other hand, Fig. \ref{fig:1}.c reveals that the average clustering 
coefficient $CC(k)$ of nodes with degree
$k$, strongly depends on $k$. In particular we observe a power law decay 
$CC(k)\sim k^{-\alpha}$ for high values of the mean clustering coefficient of the
network. Therefore, all the networks used in this work have the same 
degree distribution and no degree-degree correlations and thus they allow 
us to make a correct estimate of the role of the clustering coefficient on the 
promotion of cooperation in SF networks.

%% Dynamics
{\em Evolutionary Dynamics.-} We now assume that each node of the
graph represents a player. A link between two nodes of the graph
indicates that the two players interact and can play. We implement the
finite population analogue of replicator dynamics \cite{sp05,gcfm07}
for the PD game with payoffs $R=1$, $P=S=0$, and $T=b>1$.  At each
generation, of the discrete evolutionary time, $t$, each agent $i$
plays once with every agent in its neighborhood and accumulates the
obtained payoffs, $P_i$. Then all the players update synchronously their 
strategies by the following rules. Each individual $i$ chooses  
at random a neighbor, $j$, and compares its payoff  $P_i$ with 
$P_j$. If $P_i \ge P_j$, player $i$ keeps the same strategy for the next 
generation. On the other hand, if $P_j > P_i$, the player $i$ 
adopts the strategy of its neighbor $j$  with probability $\Pi_{i\rightarrow
  j}=\beta(P_j-P_i)$, for the next game round robin. 
Here, $\beta$ is related to the characteristic inverse time scale:  
the larger $\beta$, the faster evolution takes place. 
We assume $\beta = (\mathrm{max}\{k_i,k_j\} b)^{-1}$.  
This choice assures that $\Pi_{i\rightarrow j} < 1$ and also
slows down the invasion process from or to highly connected nodes 
\cite{sp05}. 

After a transient time, the evolutionary dynamics reaches a stationary
regime which can be characterized by the average cooperation 
index $\langle c\rangle$, defined as the overall 
fraction of time spent by all the players in the cooperator state. The 
value of $\langle c\rangle$ is computed as follows. 
After a transient time $\tau_0=5\cdot 10^3$, we further evolve the system 
over time windows of $\tau=10^3$ generations each, and we study the time 
evolution of the number of cooperators, $c(t)$. 
In each time window we compute the average value and the fluctuations of $c(t)$.  
When the fluctuations are less or equal to $1/\sqrt{N}$,  
we stop the simulation and we consider the average cooperation obtained in 
the last time window, as the asymptotic average cooperation $\langle c\rangle$ 
of the realization. In each realization we change both
the network and the initial conditions of the dynamics. All the
results reported below are averages over $10^3$ realizations for each
value of the network and game parameters ($p$ and $b$ respectively).

\begin{figure} %%%%%%%%%%%% FIG 2
\epsfig{file=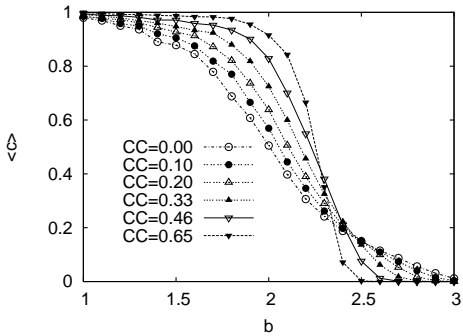,width=1.8in,angle=-90,clip=1}
\caption{Average degree of cooperation $\langle c\rangle$ as 
a function of the temptation to defect $b$. The six different 
curves show the transition from all-cooperator to all-defector states 
for SF networks with different average clustering coefficient. On the one hand,  
the cooperation is enhanced as the clustering coefficient increases. On the other hand, 
the transition to all-defector networks is smoother when clustering is smaller.}
\label{fig:2}
\end{figure}

{\em Results.-} To unveil the influence that clustering has on the
promotion of cooperation in scale-free networks, we explore the
evolutionary dynamics on networks with different values of the
clustering coefficient.  In Fig.~\ref{fig:2} we report $\langle
c\rangle$ as a function of $b$. As expected, the degree of cooperation
$\langle c\rangle$ decreases monotonously as the temptation to defect
$b$ increases. However, the path from an all-cooperator network, at
$b=1$, to an all-defector network, for high values of $b$, depends
strongly on the clustering coefficient of the SF network. From the
figure it is clear that SF networks with the highest clustering
coefficient show a remarkable survival of cooperation with values
$\langle c\rangle\simeq 1$ up to temptation values of $b=2$, in agreement 
with \cite{PRSB}. This is in contrast with the constant decrease of the cooperation observed in
networks with no clustering. On the other hand, the enhancement 
of cooperation for clustered SF networks disappears when moving 
to higher values of $b$. In particular, a sharp decrease from high to zero 
cooperation is observed when $b$ varies in the narrow range $b\in (2,2.5)$, with 
SF networks with small custering coefficients showing a slower convergence 
to the all-defector state.

%We now study the dependence of the survival of cooperation on the clustering
%coefficient from the point of view of the microscopic organization of
%cooperator strategies. 

Since all the networks analyzed share the same degree
distribution, it is possible to compare the microscopic evolution of
cooperation as a function of $b$ by looking at the probability,
$P_{c}(k)$, that a node of degree $k$ acts as cooperator in the
stationary regime. Such a probability is calculated by considering the
final time configurations for each value of $b$ and $p$.  Namely, for
a given realization $l$ (of the network and of the initial
conditions), we measure the final number $c_{l}(k)$ of cooperators of
degree $k$, and the number of nodes $n_{l}(k)$ of degree $k$. Then,
$P_{c}(k)$ is computed as $P_{c}(k)=\sum_{l}c_{l}(k)/\sum_{l}n_{l}(k)$. 

\begin{figure} %%%%%%%%%%% FIG. 3
\epsfig{file=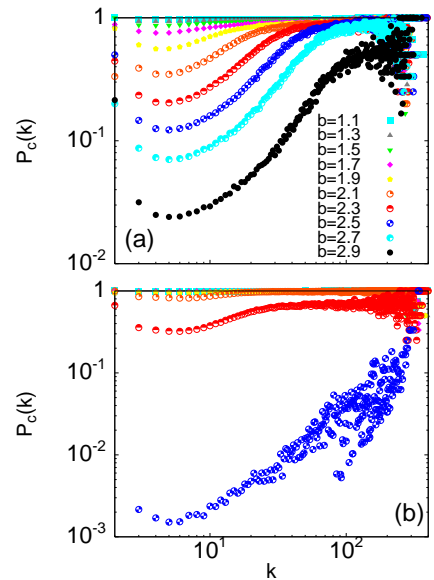,width=2.3in,angle=-0,clip=1}
\caption{Probability $P_{c}(k)$ of finding a node of degree $k$ 
playing as cooperator in the stationary regime of the evolutionary 
dynamics. Different curves correspond to different values of the 
temptation to defect, $b$. The two panels correpond to two SF with different $CC$, 
namely {\bf (a)} $CC=0.0$, {\bf (b)} $CC=0.65$.}
\label{fig:3}
\end{figure}

In Fig.~\ref{fig:3} we report $P_{c}(k)$ for different values of the temptation, $b$, and 
for two SF networks corresponding to the lowest and highest values of the $CC$. 
As $b$ increases, and hence the average cooperation $\langle c \rangle$ 
decreases, the curves $P_{c}(k)$ show the same behavior in the two networks 
considered. In particular, high degree nodes are more resistant to defection and 
display the highest values of $P_{c}(k)$, for each value of $b$. In addition to this, 
the profile of $P_{c}(k)$ shows, for all the curves, a well defined global minimum
for intermediate degree classes. Therefore, low connected nodes are not the easiest 
ones to be invaded by defectors. 
This result has been previously reported for BA networks in  \cite{Poncela}. In BA networks 
the existence of the minimum is explained by the presence of low degree nodes (the last 
nodes to be attached in the network growth process) that are only connected to the hubs. 
These leaves are thus isolated by hubs from the rest of the network and therefore imitate 
and fixate the cooperative strategy adopted by their corresponding neighboring hubs. The 
same picture applies for highly clustered networks but with an important difference regarding 
the organization of leaves around hubs. In this case, the last nodes attached 
to the network are usually connected both with a hub and with other low degree 
nodes (also attached to the hub). These nodes are again dynamically isolated 
from the rest of the network by the hub and thus they imitate and then fixate the hub's strategy. 
Additionally, the links between isolated leaves that close the triads (composed of a hub and two leaves) 
nourish these leaves with a new mechanism to resist defection since their payoff  is now provided both 
from the hubs and other leaves. Therefore, any eventual change of the state of the hubs is not trivially 
followed by a change of leaves' state since they can still obtain payoff from the interactions that share 
among them. In other words, the density of triangles around hubs in highly clustered SF networks enhances 
the fixation of cooperation in low degree nodes. 

%Therefore the existence of a minimum in the curve
%$P_{c}(k)$ at intermediate degree classes (regardless of the value of
%the CC) can be intrepreted an effect of the SF architecture. 

\begin{figure} %%%%%%%%%%% FIG. 4
\epsfig{file=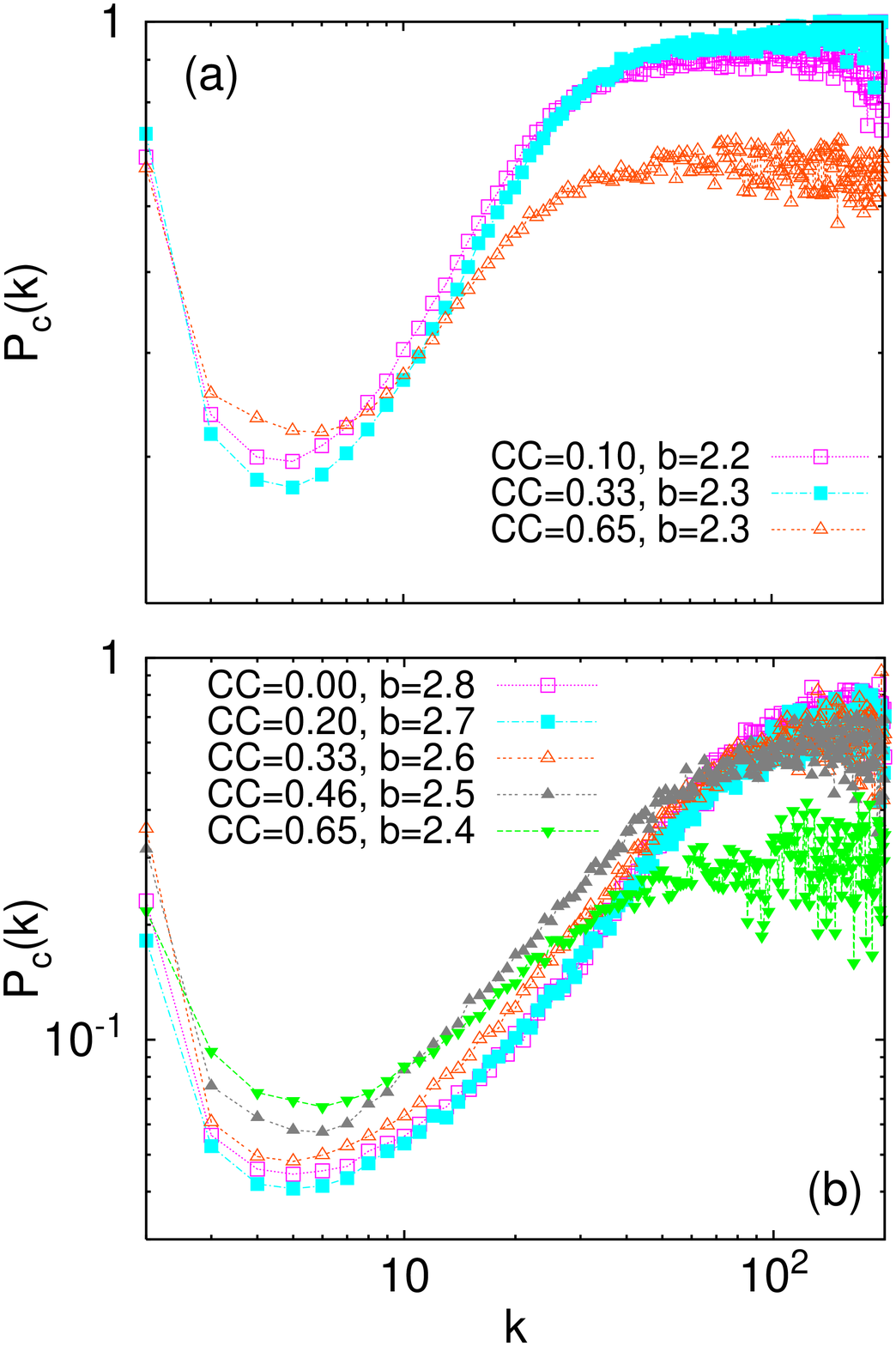,width=2.3in,angle=-0,clip=1}
\caption{Probability $P_{c}(k)$ of finding a node of degree $k$ 
playing as cooperator in the stationary regime of the evolutionary 
dynamics. Each panel shows $P_{c}(k)$ for networks with different 
$CC$ and the same average level of cooperation: 
{\bf (a)} $\langle c\rangle=0.35$,  {\bf (b)} $\langle c\rangle=0.05$.
Note that in each panel the curves $P_{c}(k)$ correspond to different 
values of the temptation to defect, $b$, for each network.
}
\label{fig:4}
\end{figure}

Let us now focus on the path towards $\langle c\rangle= 0$ as $b$ increases. Although the 
overall picture revealed from Fig.~\ref{fig:3} seems to be qualitatively the same regardless the 
$CC$ of the networks, a careful inspection of the results reveals that a high $CC$ 
tends to homogenize the role of degree classes when defectors invade the network.
In Fig. \ref{fig:4} we have plotted again the probability $P_c(k)$ 
for several SF networks of different $CC$. In each panel of the figure we have plotted 
the curves  $P_c(k)$ of several networks at different temptation values, $b$, so that the average 
level of cooperation, $\langle c\rangle$, is the same in all the networks. Namely, Figs. \ref{fig:4}.a 
and \ref{fig:4}.b correspond to $\langle c\rangle\simeq 0.35$ and $0.05$ respectively.
For low clustering networks the shape of $P_{c}(k)$  can be naively described by defining a 
quantity $k^{*}(b)$, so that for $k>k^{*}(b)$ we have $P_{c}(k)\simeq 1$, while $P_{c}(k)\ll1$  
for $k<k^{*}(b)$. This description has been already introduced in \cite{Poncela} for  BA networks. 
Obviously,  the value $k^{*}(b)$ grows with $b$ (see Fig.\ref{fig:3}.a) and hence the conversion of cooperator 
into defector strategies can be explained as a progressive invasion of the degree classes by defectors:
the larger the value of $b$ the more degree hierarchies defectors have invaded. This 
evolution points out a smooth transition towards $\langle 
c\rangle= 0$ for SF networks with low $CC$ values, as reported in Fig.~\ref{fig:2}. 
Conversely, for highly clustered SF networks there is not such critical threshold 
$k^{*}(b)$ and the invasion by defectors affects homogeneously the degree classes. 
This is clear from Figs. ~\ref{fig:4}.a and~\ref{fig:4}.b by looking at the curves 
$P_{c}(k)$ corresponding to SF networks with $CC=0.65$. In these two curves, 
corresponding to $\langle c\rangle=0.35$ and $0.05$, all the degree classes 
have been already affected by the invasion of defectors. 
%In particular, nodes with $k>20$ 
%have a probability of being cooperator  $P_c\simeq 0.6$ instead of $1$. 
%This is not observed 
%for unclustered networka where there is always a fraction of nodes (those with highest value of $k$) 
%with $P_c=1 $ whenever  the average fraction of cooperator is larger than $0$. 
%
Therefore, one cannot describe the path towards $\langle c\rangle= 0$ in highly clustered 
SF networks as a hierarchical invasion of defectors such as in the BA case \cite{Poncela}. 
On the contrary, the degree hierarchy seems not to play a crucial role 
as soon as defectors invade highly clustered networks. This result would explain the 
sudden drop of cooperation reported in Fig. \ref{fig:2} for high values of $CC$ as a consequence 
of the low ability of clustered networks to bias defector strategies towards low 
and intermediate degree classes. 
 
{\em Conclusions.-} We have studied the role of clustering, a
typical property of social systems, in the evolution of cooperation in
SF networks. Our conclusion is twofold. On the one hand, a significative 
enhancement of cooperation is shown when the clustering coefficient of the 
network is high. This enhancement is manifested by the persistence of a
population of (nearly) all cooperators in the network even for
large values of the temptation to defect. On the other hand, the transition to 
zero level of cooperation becomes sharper as the clustering of the network increases. 
%Moreover, for very large temptation values, unclustered SF networks outperforms 
%that of highly clustered ones. 
The sudden drop of the cooperation in highly clustered populations 
%has been analyzed from a microscopic point of view. Our
%results point out that the invasion of cooperator degree classes by
is explained as a consequence of the spreading of defector strategies across 
all the degree classes. Therefore, the picture of a hierarchical invasion of defectors 
previously observed in BA networks does not apply for highly clustered SF networks. 
%
%In summary, the presence of high clustering coeffcient in scale-free networks 
%enhances the stability of the all-cooperator population  but, on the other hand, 
%it homogeneizes the defector's invasion process of the degree classes, decreasing 
%the resistance of low densities of cooperator players in the network.

\begin{acknowledgments}
We acknowledge useful discussions with Y. Moreno and R. Sinatra. This work has been 
supported by INFN project TO61. 
\end{acknowledgments}

\end{document}